\documentclass[fleqn,usenatbib]{mnras}

\usepackage{newtxtext,newtxmath}

\usepackage[T1]{fontenc}

\DeclareRobustCommand{\VAN}[3]{#2}
\let\VANthebibliography\thebibliography
\def\thebibliography{\DeclareRobustCommand{\VAN}[3]{##3}\VANthebibliography}

\usepackage{graphicx}	
\usepackage{amsmath}	
\usepackage{amssymb}	

\title[]{Bi-directional streaming of particles accelerated at the {\it STEREO}-A shock on 9$^{th}$ March 2008 } 

\author[F. Fraschetti \& J. Giacalone]{
F. Fraschetti,$^{1,2}$\thanks{E-mail: ffrasche@lpl.arizona.edu}
J. Giacalone,$^{1}$
\\
$^{1}$Dept. of Planetary Sciences-Lunar and Planetary Laboratory, University of Arizona, Tucson, AZ, 85721, USA\\
$^{2}$Center for Astrophysics $|$ Harvard \& Smithsonian, Cambridge, MA, 02138, USA\\
}

\date{Accepted XXX. Received YYY; in original form ZZZ}

\pubyear{2015}

\begin{document}
\label{firstpage}
\pagerange{\pageref{firstpage}--\pageref{lastpage}}
\maketitle

\begin{abstract}
We present an interpretation of anisotropy and intensity of supra-thermal ions near a fast quasi-perpendicular reverse shock measured by {\it Solar Terrestrial Relations Observatory Ahead (ST-A)} on 2008 March $9^{th}$. The measured intensity profiles of the supra-thermal particles exhibit an enhancement, or ''spike'', at the time of the shock arrival and pitch-angle anisotropies before the shock arrival are bi-modal, jointly suggesting trapping of near-scatter-free ions along magnetic field lines that intersect the shock at two locations. We run test-particle simulations with  pre-existing upstream magnetostatic fluctuations advected across the shock. The measured bi-modal upstream anisotropy, the nearly field-aligned anisotropies up to $\sim 15$ minutes upstream of the shock, as well as the ''pancake-like'' anisotropies up to $\sim 10$ minutes downstream of the shock are well reproduced by the simulations. These results, in agreement with earlier works, suggest a dominant role of the large-scale structure ($100$s of supra-thermal proton gyroradii) of the magnetic field in forging the early-on particle acceleration at shocks.
\end{abstract}

\begin{keywords}
acceleration of particles -- shock waves -- turbulence
\end{keywords}



\section{Introduction}

The association of energetic charged particles with interplanetary (hereafter IP) shocks is an observational and theoretical pillar of space physics and astrophysics \citep[e.g.,][]{Treumann:09,Burgess.etal:12}. Diffusive shock acceleration theory \citep{Axford.etal:77,Bell:78a,Blandford.Ostriker:78,  Krymskii:77,Jokipii:82} is generally applicable only to particles at sufficiently high energy that the pitch-angle distribution (hereafter PAD) is nearly isotropic. Recent research has been vastly focusing on the origin of these so-called ``seed-particles'', that include 1) Solar Wind (hereafter SW) particles accelerated directly from the thermal pool during solar energetic particle events, likely at shocks \citep{Neergaard-Parker.Zank:12,Neergaard-Parker.etal:14,Giacalone:17}; 2) pre-existing quiescent population of supra-thermal particles, suggested early-on, e.g., by a He/H intensity ratio of $\sim 30$ keV ions upstream of the Earth bow shock correlated to the He/H intensity ratio in the SW in contrast with the intensity ratio in the field-aligned beams \citep[{\it ISEE 1}]{Ipavich.etal:84,Ipavich.etal:88}, or by the intensity of $\sim 35$ keV ions upstream of IP shocks detectable only in a small fraction of the analysed sample (3 events out of 17) by \cite{Gosling.etal:84} in {\it ISEE 1 and ISEE 3} or other works \citep{Desai.etal:06a,Desai.etal:06b,Mewaldt.etal:07a,Kahler.etal:09,Mason.etal:12b}. 
However, a mixture of both populations has not been observationally ruled out \citep{Desai.Giacalone:16}.

Multiple spacecraft (hereafter s/c) measurements at 1 AU \citep[{\it ACE}, {\it Wind},][]{Lario.etal:19} have shown in the supra-thermal energy range ($\sim 10$ keV ions in the s/c frame) a below-background spectrum upstream of some IP shocks of various magnetic obliquity, i.e., both close to parallel and to perpendicular; only very close ($\lesssim 5$ minutes) to the shock arrival a supra-thermal component emerges above the $\sim 1$ count/sec level. These observations suggest that, at least in these cases, the pre-existing suprathermal population does not play a significant role in the generation of the downstream high-energy tail of the momentum distribution. In the supra-thermal range a weak anisotropy (and low intensity) is measured within $\sim 2$ hours before the arrival of highly-oblique shocks whereas strong anisotropy is found for the low-obliquity ones, due to ions streaming nearly-aligned to the magnetic field \citep[][]{Lario.etal:19}. 
In this context an investigation of the high-obliquity shocks with a measured large anisotropy might help shed light on the supra-thermal ions dynamics.

Localised enhancements at shocks, or shock spikes, have been identified in the past decades by a number of {\it in-situ} measurements via {\it Explorer 33-35} \citep{Armstrong.etal:70} and {\it Vela 4} \citep{Singer.Montgomery:71} in the supra-thermal ($10$s, $100$s keV) ion intensity at the shock with duration of a few tens of seconds, $\sim$minutes \citep[{\it ACE}, ][]{Lario.etal:03}. The {\it Voyager-1} crossing of the solar termination shock \citep{Decker.etal:05} suggests a spike in high-energy protons ($3.4 -17.6$ MeV) and ions ($40 -53$ keV) intensities \citep{leRoux.etal:07,Zuo.Feng:13}.  The measured anisotropy makes spikes good candidate to improve our data-driven understanding of the role of the large scale magnetic geometry in the early-on particle acceleration. 

Formerly postulated to originate from ions anisotropically reflected between  Earth bow and IP shocks \citep{Axford.Reid:63}, spikes were first modelled by \citet{Decker:83} for quasi-perpendicular shocks ($\theta_{Bn} > 70^{\rm o}$, 
where $\theta_{Bn}$ is the angle between the local shock normal and the upstream unperturbed magnetic field) by using shock drift acceleration of a seed particle population. An alternative scenario of spikes originated as ions are reflected by the magnetic barrier and conserve the first adiabatic invariant applies to the quasi-perpendicular case only \citep{Gieseler.etal:99}. Enhancements associated with $\theta_{Bn} < 70^{\rm o}$ need an additional source of turbulence arguably provided by the self-excited waves produced by energetic ions streaming upstream \citep{Scholer:85}.     

\cite{Erdos.Balogh:94} proposed that spikes are generated by magnetic trapping of particles formed by multiplecrossings of the turbulent magnetic field lines with the shock surface; a supra-thermal ion beam aligned with the upstream field would produce a single-peak PAD within the loss-cone.  
\cite{Erdos.Balogh:94} showed qualitative agreement between the {\it ISEE-3} data in the $5$ minutes preceding the IP shock on 1978 Dec $25^{th}$ and a numerically calculated double-peak PAD. Such a double peak results from a double loss cone, footprint of a bi-directional streaming along the upstream field crossing the shock surface in multiple points. A subsequent analysis of {\it Ulysses} data \citep{Marhavilas.etal:03} of the spike at the quasi-perpendicular ($\theta_{Bn} \gtrsim 75.1^{\rm o}$) shock event 1992, DOY256, 
shows a clear bimodal PAD (see Fig. 4, a, b therein) and estimates the length scale of the trapping region $\gtrsim 10$ times the supra-thermal ions gyroradius. 

The present paper is devoted to the interpretation of the measured PADs and intensity of suprathermal ions in a recent {\it STEREO}/A ({\it STA}) quasi-perpendicular ($\theta_{Bn} \simeq 71^\circ$) reverse shock of a stream interaction region (2008, DOY 069, March $9^{th}$, UT: 19:50); such an event was found to exhibit a bimodal PAD in the supra-thermal range ($10-40/$nuc keV H$^+$ and $8-20/$nuc keV He$^+$)  in the $6$ minutes upstream of the shock \citep[see Fig. 5(e,f) in ][]{Yang.etal:20} and transiting to a single-peak shape aligned with reflected ions far upstream of the shock. A pancake-like PAD was measured downstream. 

{\it In-situ} measurements of SW thermal density and magnetic field have long shown that turbulence is an inherent property of the heliospheric plasma from sub-ion scale up to the correlation length \citep[$\sim 0,01$ AU at $1$ AU, e.g.,][]{Jokipii.Coleman:68,Bruno.Carbone:05}. By implementing a pre-existing magnetic turbulence frozen with the fluid and passively advected through the shock  \citep{Decker.Vlahos:86,Giacalone.Jokipii:09}, test particle numerical simulations were used \citep{Fraschetti.Giacalone:15} to determine the PAD of supra-thermal ions as the shock approaches the s/c. In particular, the emergence of a bi-modal PAD close to the shock from a reflected ions-dominated regime far upstream was investigated in detail.

In this letter, we use the analysis of the {\it STA} reverse shock on 2008 DOY 069 in \cite{Yang.etal:20} to reproduce with test-particle simulations for the first time the supra-thermal protons PADs at a shock in quantitative agreement with {\it in-situ} measurements. 

\section{{\it STEREO} observations}\label{sec:data}

The {\it STA} 2008 DOY069 (March $9^{th}$, UT: 19:50) shock event was observed within a stream interaction region. {\it In-situ} measurements constrain the local geometry of the shock and the local magnetic field orientation. The measured shock parameters used here  are\footnote{As reported in the archive http://ipshocks.fi/database}: density compression across the shock $r=2.01 \pm 0.30$, $\theta_{Bn}=71.5^\circ \pm 20^\circ$,  upstream total magnetic field $B = 9.45 \pm 0.46$ nT, upstream Alfv\'en speed $V_A = 82 \pm 8$ km$/$s, velocity of the shock along the shock normal in the s/c frame $V_{sh} = 237 \pm 173$ km$/$s, SW speed in upstream frame $V_u \sim 662$ km$/$s, velocity of the shock in the upstream frame along normal $U_1^x =|V_{sh} + {\bf V_u} \cdot {\bf n}| \simeq 147.6$ km$/$s, where $+ {\bf V_u} \cdot {\bf n}$ follows the sign convention in the archive above and indicates the SW speed projection along the shock normal in RTN coordinates with ${\bf n} = [-0.64 \pm 0.22, -0.76 \pm  0.20, 0.11 \pm 0.30]$, directed sunward for a reverse shock and the $x$-axis is directed along the average shock motion, Alfv\'en Mach number $M_A = U_1^x/V_A = 1.8 \pm 0.7$, sonic Mach number $M_s=1.3 \pm 0.5$. \cite{Yang.etal:20} define three time  intervals across the {\it STA} measurements ($U1 \sim 6$ minutes upstream of the shock, $U2 \sim 10$ minutes upstream of $U1$ and $D1 \sim 10$ minutes downstream of the shock)\footnote{For the time-intervals $U1$ and $U2$, the notation in \cite{Yang.etal:20} is adopted hereafter, not to be confused with the plasma speeds in the shock frame, customarily labelled as $U_1$ and $U_2$.} and provide separate PADs within each interval. During $U2$ the magnitude of $B$ decays \citep{Yang.etal:20} from the forementioned upstream value $B= 9.45$ nT (measured within $\sim 10$ minutes from the shock) to the typical value at $1$ AU ($B \sim 5$ nT). Simulations of particle motion as far upstream as $U2$ should account for the slow decompression of the unperturbed field therein. Here the value $B \sim 5$ nT is adopted.  

\section{Numerical set-up}\label{sec:numericalsetup}

\subsection{Synthesized turbulence}\label{sec:turbulence}

\begin{figure}
\includegraphics[width=\columnwidth]{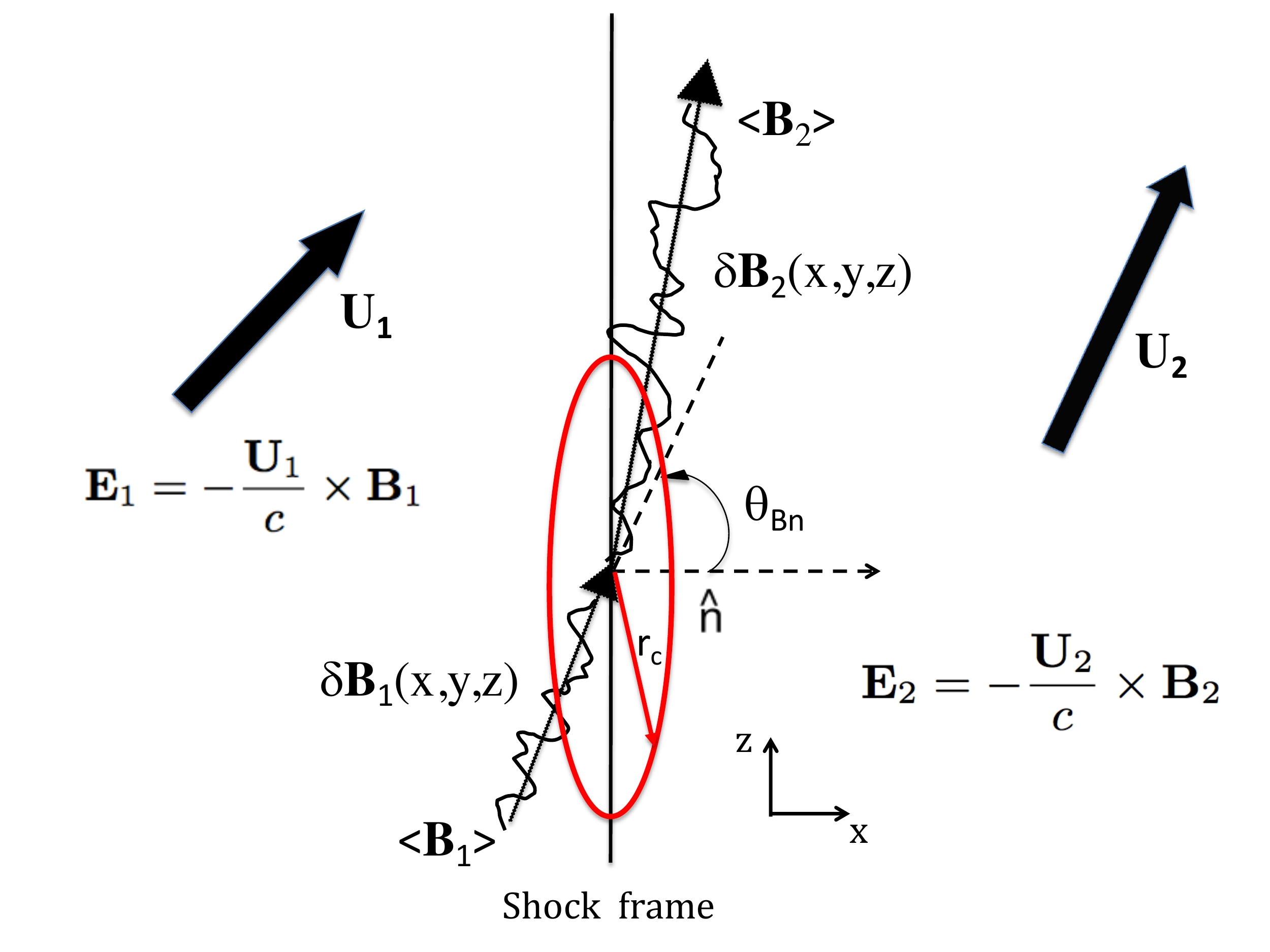}
\caption{Configuration of magnetohydrodynamic variables  \citep[adapted from][]{Giacalone:05a, Fraschetti.Giacalone:15} for the {\it STA} 2008 DOY 069 shock event (see Sect. \ref{sec:data}). The red circle of radius $r_c$ marks the effective area of the supra-thermal particles detectors (see Sect. \ref{sec:finite_extent}).}
\label{fig:cartoon}
\end{figure}
We carry out test-particle numerical simulations of supra-thermal protons at a fast, planar collisionless shock travelling in a plasma with an embedded turbulent magnetic field. The upstream three-dimensional magnetic field is given by ${\bf B(x) = B}_0 + \delta {\bf B(x)}$, with an average component ${\bf B}_0 $ having orientation $\theta_{Bn}$ with respect to the local shock normal and a random component  ${\bf \delta B} = {\bf \delta B} (x, y, z)$ having a zero ensemble average ($\langle \delta {\bf B(x)} \rangle = 0$); the turbulence is synthesized as superposition of plane waves with random amplitude, phase and orientation, with amplitudes determined by an assumed power spectrum  \citep[c.f.~][]{Decker.Vlahos:86,Giacalone.Jokipii:99,Fraschetti.Giacalone:12}. Simulations aiming at reproducing measurements of a single shock event can make use of one specific turbulence realization, rather than an ensemble average. We fit the PADs of H$^+$ within the time intervals $U1$, $U2$, $D1$, provided that the particle intensity satisfactorily reproduces the measured shape. 

For the sake of simplicity, we assume the upstream three-dimensional magnetic turbulence to be isotropic and scale-invariant, with a Kolmogorov power spectrum, in the inertial range $[k_0, k_{max}]$, where $k_0 = 2 \, \pi/L_c$, $L_c$ is the correlation length of the  turbulence \citep{Giacalone:05a,Fraschetti.Giacalone:15}, $k_{max} = 2 \, \pi/L_{min}$ and $L_{min} =  10^{-2} \, L_c$. At smaller wavenumbers, or larger scales, in the range $[k_{min}, k_0]$, where $k_{min} = 2 \, \pi/L_{max}$ and $L_{max} =  10^{2} \, L_c$, the power spectrum is assumed to be uniform. Such a power spectrum includes only the pre-existing SW fluctuations and not those ``self-generated'' by the energetic ions streaming ahead of the shock via, e.g.,  cyclotron-resonant streaming instability \citep{Tademaru:69,Lee:83}. 
In the simulations presented here, we set the unperturbed upstream magnetic field $B_0 = 5$ nT (see Sect. \ref{sec:data}) and $L_c = 0.01$ AU, so that $r_g/L_c \ll 1$, where $r_g$ is the particle gyroradius, at all particle energies considered here and the condition of resonant scattering with all turbulent wavenumbers is satisfied. The dataset for the {\it STA} 2008 DOY069 event  (see Sect. \ref{sec:data}) restricts the simulations parameter space to the power of the magnetic turbulence ($\delta B^2$) relative to the power of the  unperturbed field ($\sigma^2 = \delta B^2/B_0^2$). We have verified that a turbulence power spectrum extended to higher wavenumbers, or smaller scales, from $10^2 \, k_0$ up to $10^3 \, k_0$ does not alter significantly the bi-modal shape of the PAD, nor the particle intensity.

The geometrical configuration implemented in the simulations is summarized in Fig.  \ref{fig:cartoon}. The plane $x=0$ marks the planar shock surface. Bulk plasma flows into the shock from $x<0$ to $x>0$ 
along the shock normal, along the x-axis. The components of the bulk flow velocity before and after the shock satisfy the jump conditions at an infinitely planar shock. The profile of the bulk velocity along the x-axis is taken to be discontinuous to the scale of the supra-thermal particles, much larger than the ion skin depth. The reference frame is chosen so that ${\bf U}_1$ lies on the xz-plane and ${\bf U}_1 = (U_1^x, 0, U_1^z)=(147,0, 185)$ km$/$s (see Sect. \ref{sec:data}).

We determine the trajectory of individual particles by numerically solving the Lorentz equation in the prescribed magnetic field. Assuming that in the plasma rest frame both upstream and downstream electric field vanish (infinitely conductive), particles undergo acceleration by the motional electric field that writes in the shock frame ${\bf E} (x, y, z, t) = - {{\bf U}(x) / c} \times {\bf B}(x, y, z, t) $, where ${\bf U} (x) = (U_x, 0, U_z)$ and the magnetostatic field ${\bf B}(x, y, z, t)$ is passively advected along the flow \citep{Giacalone.Jokipii:09,Fraschetti.Giacalone:15}, where the $t$-dependence of ${\bf B}$ at a given location can be interpreted as a spatial variation as it results from the fluid advection only. Since the particles trajectories are integrated over a single turbulence realization, the phase-space distribution function in the time-intervals $U1$, $U2$, $D1$ represented here is not solution of a steady-state ensemble-average transport equation, as it corresponds to a particular realization of ${\bf B}$; the time variation of a chosen realization ${\bf B}$ at a given spatial location could average out with an equal and opposite time-variation at the same location of a distinct realization of ${\bf B}$ with no change for the ensemble average (see Sect. \ref{sec:disc_concl}).

The boundary conditions are based on the approach in \citet{Giacalone:05a,Fraschetti.Giacalone:15}: the motion is tracked until either {\it a)} particles reach a pre-specified high-energy cut-off, taken to be much higher than the energy of interest in this study ($p_b = 500\, p_0$ where $p_0$ is the injection momentum in the plasma frame), or {\it b)} particles escape by advection at a downstream boundary ($x_b = 2.5 \times 10^4 |{\bf U}_1|/\Omega_0$), where the prescription of the return probability is implemented \citep{Ellison.etal:96,Giacalone:05a,Fraschetti.Giacalone:15}.  
No energy-independent free-escape boundary upstream is assumed. Ions are injected upstream on the plane $x_0=-r_g^0/2$, where $r_g^0$ is the initial gyroradius, at a random location in the plane $yz$ within a square of side $L_{max}$ to capture the effect of the perpendicular transport due to the field line meandering that originates from scales larger than $L_c$ \citep[e.g.][]{Jokipii:66,Webb.etal:09,Fraschetti.Jokipii:11,Laitinen.Dalla:17} with no ignorable coordinate \citep{Jokipii.etal:93,Jones.etal:98}. Particles are injected with an isotropic distribution in the local pitch-angle whose cosine is defined as $\mu_{local} = {\bf p} \cdot {\bf B}/pB$, where ${\bf p}$ is the particle momentum in the local plasma frame.
 
\subsection{Effect of the finite extent of the shock on pitch-angle distributions}\label{sec:finite_extent}

Within each given spatial interval along $x$-axis and energy interval, our simulations enable binning all ions positions and $\mu_{local}$ that spread over an infinite area plane parallel to $x=0$. However, due to the limited effective area of the detectors onboard {\it STA}, compared to the shock surface, such a simulation set-up introduces a bias in the PAD with respect to the s/c measurements: particles moving nearly along the shock surface leave the volume spanned by the detector during a given time interval $\Delta t$ faster than particles nearly aligned with the shock normal and, due to the finite extent of the shock, the former are not replenished as fast as the latter. As a result, in the quasi-perpendicular configuration of the {\it STA} 2008-069 event, the measured intensity of field-aligned particles ($\mu \sim \pm 1$ or $0^\circ - 180^\circ$ pitch-angle) is depleted compared to the intensity of those at $\mu \sim 0$ ($90^\circ$ pitch-angle); such an effect has to be accounted for in the data-modelling. In addition, the effective area of {\it PLASTIC} (Plasma and Suprathermal Ion Composition) instrument in the supra-thermal ions energy band ($10-40$ keV$/$nuc) is comparable with the one of {\it SEPT} (Solar Electron and Proton Telescope) instrument in the energy band $\sim 80-350$ keV$/$nuc\footnote{For ions the {\it PLASTIC} geometric factor is $0.1$ cm$^2 /$sr \citep[Table 1 in][]{Galvin.etal:08}, for {\it SEPT} it is $0.17$ cm$^2 /$sr \citep[Table 4 in][]{MuellerMellin.etal:08}.}. Thus, both the {\it PLASTIC} and {\it SEPT} geometric factors, proxy for the effective collecting area of particles, are much smaller than the physical size of the detector.  
We account for this instrumental limitation by introducing the {\it PLASTIC/SEPT} collecting area as an additional parameter as illustrated below. We also note that {\it PLASTIC} points ``nearly''-sunward whereas the {\it SEPT} ``Sun'' telescope points  $\sim 45^\circ$ from the Sun-{\it STA} direction (along the nominal Parker spiral). Thus, the viewing angles of {\it PLASTIC} and {\it SEPT} are different.

In the numerical simulations, a collecting area as small as the effective detector area requires a very large number of test-particles due to the low density of suprathermal particles.  
For the sake of simplicity, we use an energy-independent effective collecting area 
assimilated to a circle in the $y-z$ plane of radius $r_c = A\, L_c$, where $A$ is a constant to be determined by the combined fit of the PADs in the three time intervals $U1$, $U2$ and $D1$ (see Fig. \ref{fig:cartoon}). The motivation for the proportionality of $r_c$ to $L_c$ is that, by definition of $L_c$, the transport of particles within a distance $L_c$ from the detector is governed by a  cascading turbulence correlated with the detector location. Scales $> L_c$ are expected to contribute the PAD as the perpendicular transport is governed by magnetic fluctuations at such scales  \citep[field-line meandering, e.g.][]{Jokipii:66,Qin.etal:02, Minnie.etal:09,Fraschetti.Jokipii:11}. The PADs in \cite{Yang.etal:20} cover a total time interval $\Delta t \simeq 26$ min, including $\Delta t_{U1} \simeq 6$ min, $\Delta t_{U2} \simeq 10$ min and $\Delta t_{D1} \simeq 10$ min. For comparison, the scatter-free travel time of the highest {\it PLASTIC} energy particles is $L_c/v(40{\rm keV}/{\rm nuc}) \sim 8.9$ min $< \Delta t$. Thus,  as the shock travels, particles originating from a shock region at a distance $> L_c$ are detectable. 
However, particles are injected within a square of side $L_{max}$ on the plane $x=x_0$, very close to the shock surface. Therefore, the range of $A$ to be explored spans from $1$ to $L_{max}/L_c = 10^2$.  Due to the lack of further constraints, ${\it A}$ has to be determined directly by comparing simulations with data.

\section{Results}\label{sec:results}

We calculated PAD and intensity profiles for suprathermal protons in the {\it PLASTIC} and {\it SEPT} energy bands, respectively, for $\sigma^2= 0.1, 0.3. 0.4, 0.5, 0.7, 1.0, 1.2$ for a number of magnetic turbulence realizations in each case. Given the {\it PLASTIC} supra-thermal energy band for H$^+$ ($10 - 40$ keV$/$nuc), the goldilock injection particle energy in the local plasma frame, $E_0$, has to be lower than the low {\it PLASTIC} energy bound, $10$ keV, to enable all particles to undergo shock-acceleration and resonantly interact with the upstream turbulence before accessing the {\it PLASTIC} energy band. In addition, the particle injection speed $v_0$ (in the upstream plasma frame) has to satisfy $v_0/U_1^x \lesssim 10$ as higher speed particles quickly isotropize downstream, as predicted by DSA. 
The value $E_0 = 3$ keV ($k_0 \, r_g^0 \simeq 0.67$) satisfies such conditions and will be used hereafter.

\begin{figure}
\hspace{-0.6cm}
\includegraphics[width=9.2cm]{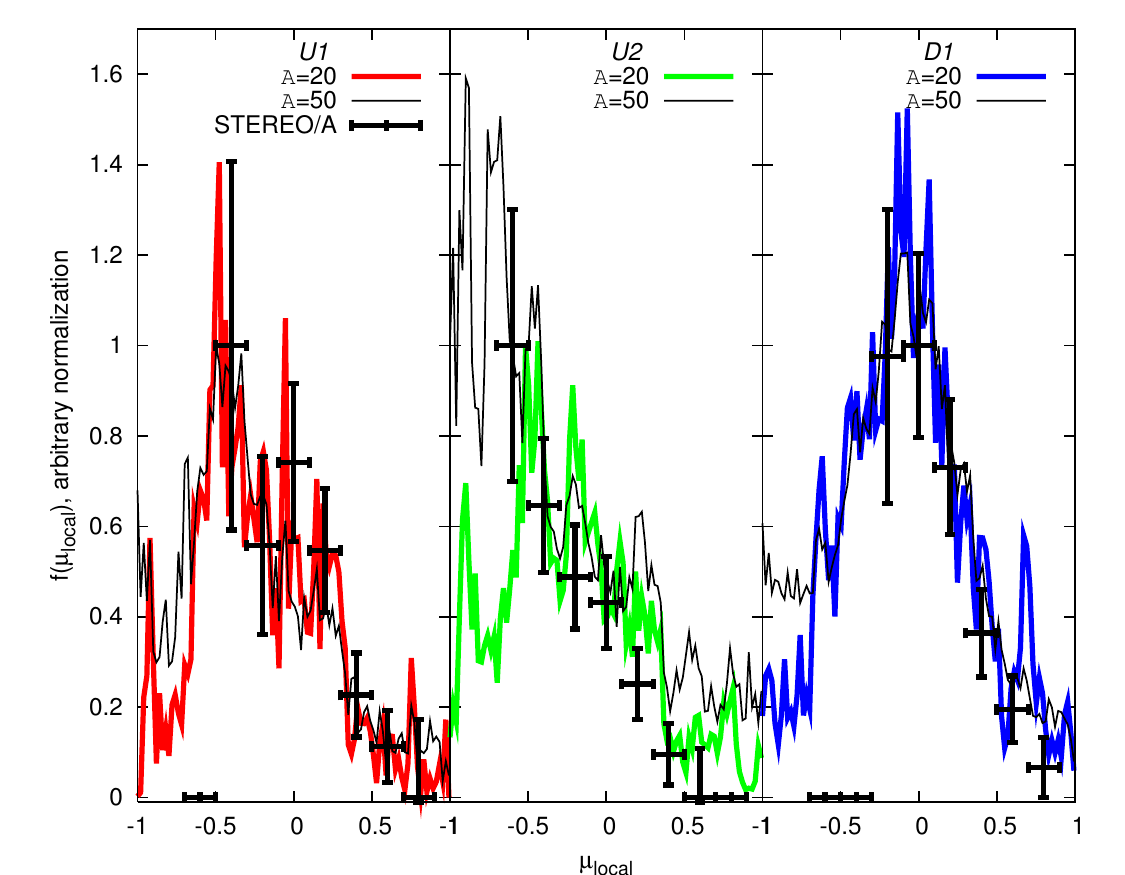}
\caption{Numerical PADs of $10 - 40$ keV protons for the time intervals {\it U1} (left panel), {\it U2} (middle panel) and {\it D1} (right panel) for $\sigma^2=0.3$ and for distinct values of ${\it A} =20$ (red for {\it U1}, green for {\it U2} and blue for {\it D1}) and ${\it A} = 50$ (black in each panel) compared with {\it STA/PLASTIC} data \citep{Yang.etal:20} for $H^+$ in the same energy range.
}
\label{fig:PAD_var_x_cyl}
\end{figure}

Figure \ref{fig:PAD_var_x_cyl} compares the PADs during {\it U1, U2} and {\it D1} for distinct values of ${\it A} < L_{max}/L_c = 10^2$ with {\it STA} data. The expected and measured underlying monotonic upstream PAD (decreasing from $\mu <0$ to $\mu >0$) due to reflected ions is found both during {\it U1} and {\it U2}. The single realization of $\mathbf{\delta B}$ shown here remarkably reproduces the bi-modal structure in {\it U1} (Fig. \ref{fig:PAD_var_x_cyl}, left panel, red curve)  peaking at $\mu \sim -0.4$ and $\mu \sim 0.0$ with ${\it A} = 20$, due to particles streaming in low-scattering regime along the magnetic field lines \citep{Erdos.Balogh:94, Marhavilas.etal:03,Fraschetti.Giacalone:15}. The peak $\mu\sim 0.0$ is smeared out in the case of ${\it A} = 50$ (black curve), possibly due to broader range of angles between the local $\bf B$ and the local shock normal, captured by a larger collecting area ({\it A}), that enables isotropization of the PAD at $\mu\sim 0.0$. We note that the fit of the PAD during {\it U1} only cannot constrain  ${\it A}$ due to the degeneracy introduced by the choice of a single turbulence realization: the combined fit of PADs in {\it U2} and {\it D1} allows one to reduce such a degeneracy with a larger degree of confidence\footnote{We note that for the {\it STA} data \citep{Yang.etal:20} $\mu$ is calculated in the SW frame, whereas the simulations use the local plasma frame; our tests show that this inertial-frame effect does not introduce an error significant to the data-fit, due to the coarse {\it PLASTIC} $\mu$-resolution.}.

As for {\it U2} (Fig. \ref{fig:PAD_var_x_cyl}, middle panel), the bi-modal structure seen in {\it U1} is smeared out during {\it U2}, and simulations with ${\it A}=20$ (green curve) well reproduce the monotonic trend; the transition from bi-modal to monotonic shape further upstream was emphasized in previous ensemble average calculations \citep{Fraschetti.Giacalone:15}. A major difference between the two cases ${\it A} = 20$ and ${\it A} = 50$ is the  intensity excess for ${\it A} = 50$ for $\mu > 0.5$ (even higher in our runs for ${\it A} \gg 10^2$) due to the increase of the detector area (Sect. \ref{sec:finite_extent}). An additional difference between the cases ${\it A} = 20$ and ${\it A} = 50$ is the high reflected ions PAD of the latter for $\mu < -0.5$. A bending of $f(\mu)$ as $\mu \rightarrow -1$ in {\it U2} is suggested by the measured vanishing of $f$ at $\mu=-0.6$ in {\it U1}, favouring the case ${\it A} = 20$; however, $f(\mu < -0.5)$ is not constrained by data during {\it U2}. Finally, a bump in the case of ${\it A} = 50$ between $\mu= 0.$ and $0.3$ stands out over an underlying monotonic shape between $\mu= -0.6$ and $1$. Such a bump might result again from the larger collecting area (larger ${\it A}$) that intersects a larger number of magnetic field lines crossing the shock with a higher chance of capturing a second stream of ions originating from a further region on the shock ({\it U2} interval extends between $\sim 6$ and $\sim 16$ minutes upstream of the shock).

\begin{figure}
\includegraphics[width=\columnwidth]{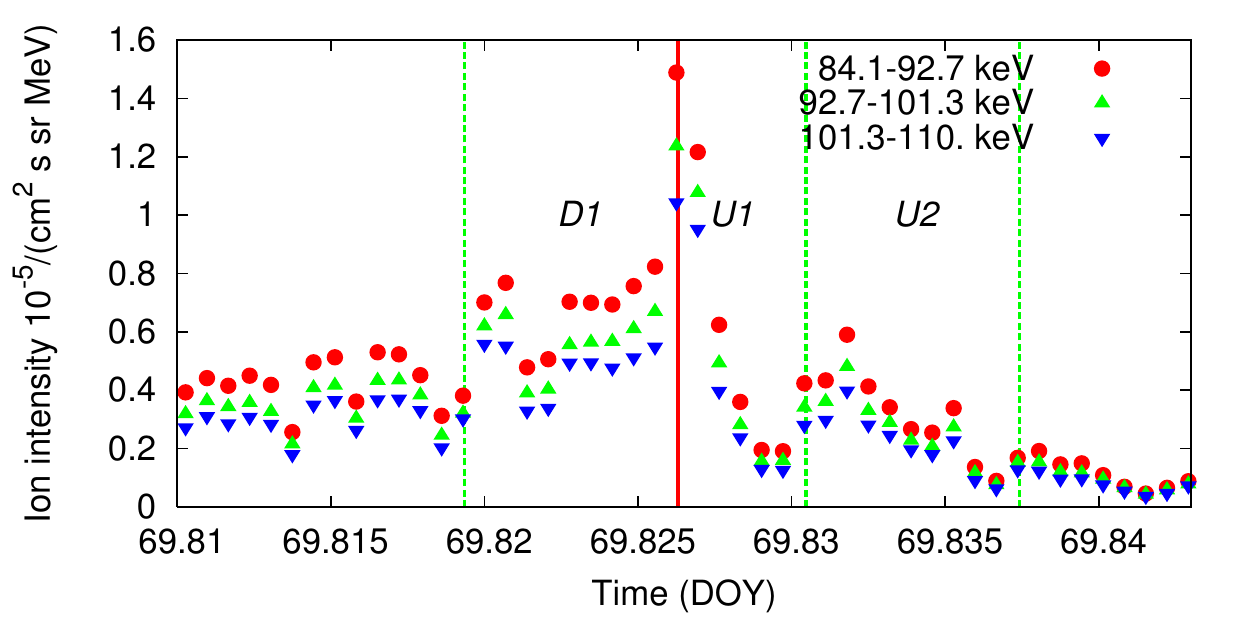}
\caption{Intensity profile from {\it STEREO} A/IMPACT/SEPT ion data, sensor pointing Anti-Sunward (level 2), 1-minute averaged in the $3$ lowest energy bands (from http://www2.physik.uni-kiel.de/stereo/index.php?doc=data). The vertical red line marks the shock passage and the green dotted lines bound the three time windows {\it U1}, {\it U2} and {\it D1}  \citep{Yang.etal:20}. A local enhancement (spike) is clearly visible at the shock in all bands.}
\label{fig:Intensity}
\end{figure}

\begin{figure}
\includegraphics[width=\columnwidth]{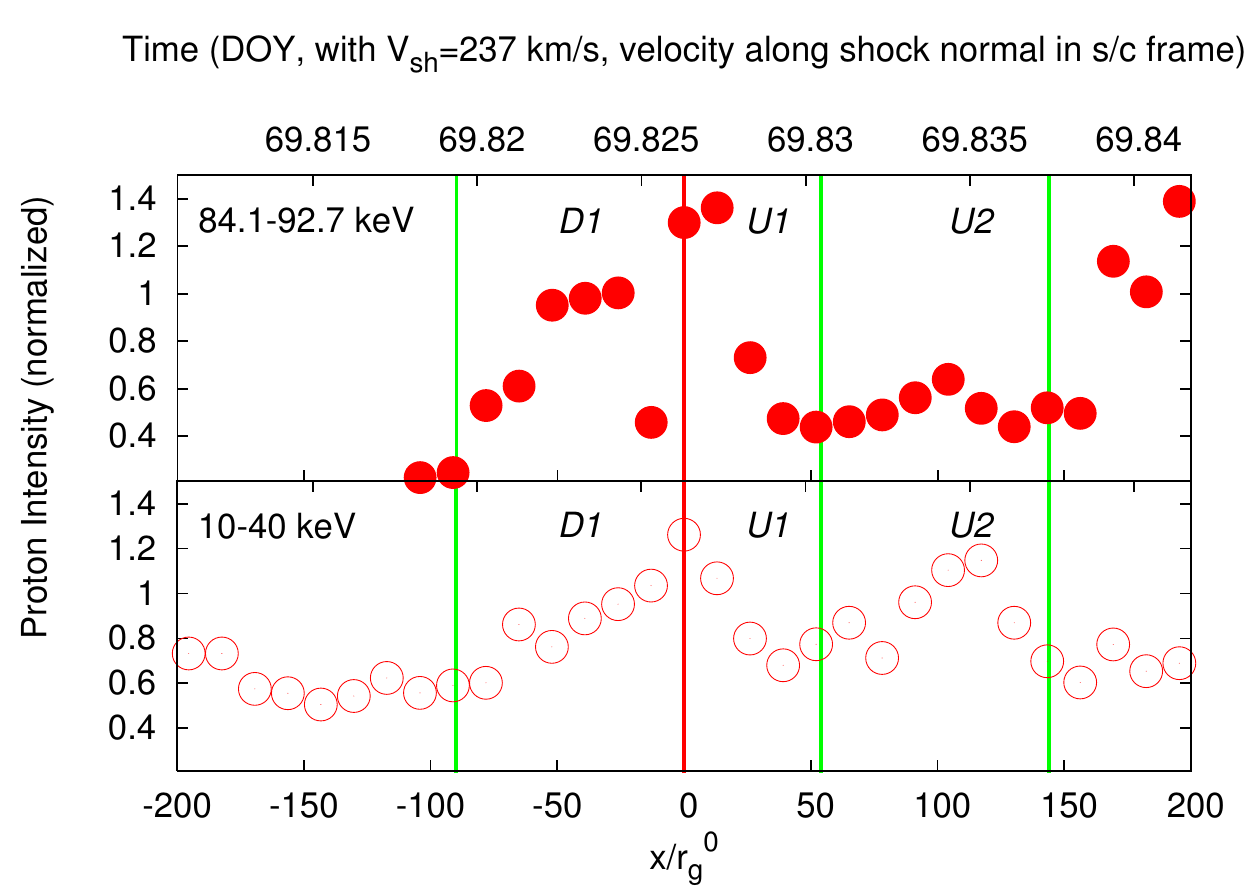}
\caption{Simulated proton intensities (lower panel) in the {\it PLASTIC} energy band $10-40$ keV, and (upper panel) in the lowest {\it SEPT} energy band (84.2 -92.7 keV, cf. Fig. \ref{fig:Intensity}) with $\sigma^2 = 0.3$ and ${\it A} = 20$. The lower x-axis is in units of injection gyroradius, the upper x-axis in units of time in the s/c frame and the y-axis is normalized to an estimated downstream intensity. The vertical red line marks the shock passage and the green dashed lines bound the three time windows as labelled in \citet{Yang.etal:20}.}
\label{fig:Intensity_simu}
\end{figure}

The right panel in Fig. \ref{fig:PAD_var_x_cyl} depicts the PAD during {\it D1}. The peak at $\mu\sim 0.0$ results from downstream advection of particles closely moving along magnetic field lines in a quasi-scatter-free scenario \citep{Erdos.Balogh:94,Fraschetti.Giacalone:15}. A difference between the cases ${\it A} = 20$ and ${\it A} = 50$ emerges only for $\mu < -0.75$, not constrained by data. 
Figure \ref{fig:Intensity} shows {\it STA}/SEPT particles intensity in the $3$ lowest suprathermal energy bands exhibiting an enhancement beyond compression at the shock arrival. During {\it U2} the intensity enhancement is well past; thus, strong deviations from monotonic PAD in {\it U2} are not to be expected. Figure \ref{fig:Intensity_simu} shows the simulated proton intensities profiles in the {\it PLASTIC} energy band $10-40$ keV (lower panel), and in the lowest {\it SEPT} energy band $84.2 -92.7$ keV (upper panel), with parameters fixed by the PADs fit in Fig. \ref{fig:PAD_var_x_cyl} ($\sigma^2 = 0.3$ and ${\it A} = 20$) and no additional fitting. An enhancement at the shock in both panels, correspondent to the spike, seems to emerge. The intensity increase in the upper panel ($84.1-92.7$ keV) ahead of {\it U2} is lacking in the observations; however, we note that the turbulence realization is chosen to reproduce PAD and intensity during {\it U1}, {\it U2} and {\it D1} intervals only. The intensity before or past such intervals is  affected by larger scale structure of the synthesized magnetic field. 

\section{Discussion and Conclusion} \label{sec:disc_concl}

The length scale $\Delta x$ of the upstream magnetic trap  originating the local intensity enhancement and the bi-modal anisotropy of $< 40$ keV/nuc protons ($r_g (40 {\rm keV}) = 5,700$ km in the upstream field) can be estimated as $\Delta x > \Delta t_{U_1} \, U_1^x = 53,000 \, {\rm km} \simeq 9.3 r_g (40 \, {\rm keV})$. For a scattering mean free path $\lambda < \Delta x \simeq 10 \, r_g (40 \, {\rm keV})$, the trap not only enables particles confinement, but also allows nearly scatter-free streaming along field lines that cross the shock in multiple points (the minimum value $\lambda = r_g$ is realized in stronger turbulence, i.e., $\sigma^2 \gtrsim 1$, whereas we present here the case $\sigma^2 \sim 0.3$).

One caveat regarding our simulations is that we have used the magnetic obliquity inferred along the s/c trajectory crossing the shock over its detector area, whereas such an angle ($\theta_{Bn}= 71.5^\circ \pm 20^\circ$ for the 2009 DOY 69 event) can fluctuate, both in the spatial location and in time, along the shock surface due to its inherent corrugation  \citep{Neugebauer.Giacalone:05,Koval.Szabo:10,Ebert.etal:16}; multi-s/c IP shocks measurements reported fluctuations $\sim 15^\circ -20 ^\circ$ over a scale $< 10^5 - 10^6$ km \citep[{\it ACE}, {\it Wind} and {\it Geotail, }][]{Terasawa.etal:05} or down to $10^3$ km  \citep[{\it Cluster}, ][]{Kajdic.etal:19}, confirmed by hybrid simulations for kinetic ions and fluid electrons \citep{Kajdic.etal:19}. Such a variability of $\theta_{Bn}$ does not alter our interpretation of magnetic trapping as a rippled shock surface propagating into a laminar medium has comparable confinement effects, as far as the ion scales are concerned, to a planar shock propagating into a turbulent field, as pre-existing turbulence corrugate the shock at scales much larger than electron-scales micro-instabilities and over longer time scales.  
Finally, the media upstream of Earth bow shock and IP shocks have been long measured to deviate from  laminar structure, since \cite{Fairfield:74}, as a result not only of ions-generated waves but also of electrons-scale kinetic effect, such as energy exchange between electrons and ions mediated by large amplitude whistlers precursors \citep{WilsonIII.etal:17}. Such scales needs kinetic ions-electrons particle-in-cell simulations and are therefore beyond the scope of this work. 

We have neglected the electric field of the order of $v_A B/c$ arising from the magnetic field fluctuations modelled as Alfv\'en waves propagating along the field at speed $v_A$. 
This is justified by the fact that for the parameters used in our test-particle simulations, $v_0 \gg v_A$ on both sides of the shock. 
Stochastic acceleration, relevant as the particles move away from the shock, is expected to have a smaller effect on the spike formation because the electric field $v_A B/c$ is disordered and has therefore an average negligible effect  \citep{Fraschetti.Giacalone:15}. 

An additional set of runs includes the PADs and intensity profiles of supra-thermal particles for a phase-space distribution function that solves the time-dependent equation of transport \citep{Giacalone:05a}; both the ensemble-averaged case and a few cases of single turbulence realization have been explored with the given turbulence power spectrum. Such cases do not provide a significantly better fit of PADs {\it U1}, {\it U2} and {\it D1} and are therefore not shown here. 

The turbulence upstream the shock is modeled herein as a superposition of plane waves with random amplitude, phase and orientation and is described by the power spectrum via normalisation ($\sigma^2$) and power law index within the inertial range ($11/3$). In the past decade the interpretation of 3D large-scale magnetohydrodynamic numerical simulations and of {\it in-situ} measurements has advocated the role of current sheets, or more generally of large gradients of the magnetic field such as discontinuities, in driving turbulence \citep{Matthaeus.etal:15}, for example, in the SW \citep{Greco.etal:16}. Dropouts in solar energetic particles count rates suggest crossing by the s/c of particle-filled flux tubes bounded by discontinuities and adjacent to particle-voided flux tubes \citep{Mazur.etal:00}; such dropouts can be measured if the flux tubes are magnetically connected with the solar photosphere and map its super-granulation  \citep{Giacalone.etal:00}, assuming a negligible cross-field diffusion. In an alternative scenario, if time dependence of perpendicular diffusion is included \citep[e.g.,][]{Fraschetti.Giacalone:12} dropouts are consistent with observations provided that the diffusion coefficient becomes large afterward \citep{Laitinen.etal:13b}. An inspection of the 1/8 second resolution magnetic field vector components/magnitude time series of the 2008 DOY069 event (IMPACT /MAG) within $16$ minutes ahead and past the shock shows no evidence of discontinuities (large fluctuations or sharp changes of direction), although a statistical analysis following, e.g., the criteria of \cite{Burlaga:69} on rotation of the B-field $> 30^\circ$ or of \cite{Tsurutani.Smith:79} on $\delta B/B > 0.5$ to identify discontinuities has not been carried out. Consistently, no impulsive voids, or dropouts, in the count rate of supra-thermal particles were observed in {\it PLASTIC} or {\it SEPT}. Thus, data suggest that the turbulence in the proximity of the shock event STA 2008, DOY 069 is not dominated by current sheet structures. Acceleration of charged particles at current sheets with strong guide field \citep[$\sigma^2 \ll 0.1$,][]{Dmitruk.etal:04,Zhdankin.etal:13} or with strong turbulent fluctuations \citep[$\sigma^2 > 1$][]{Isliker.etal:17} have been numerically analysed. In the latter case, reconnecting current sheets (as coherent structures) can mix with large amplitude magnetic fluctuations and cannot be neglected in the time evolution of the turbulence. As aforementioned, data do not show significant evidence of reconnecting current sheets. Therefore, a representation of the turbulence as plane waves superposition with an upper boundary $\sigma^2 \sim 1$ is expected to provide a realistic representation of the turbulence crossed by the STA 2008 DOY 069 shock. However, we note that, since the spacecraft probes the SW only along its trajectory, the role of current sheet in the turbulence crossed by the shock over its entire spatial extent cannot be  ruled out, even in regions of the shock magnetically connected with the s/c; however, due to the lack of evidence, it is not included herein. The interplay of current sheets and shocks in the acceleration of particles has been analyzed in several recent works \citep{Matsumoto.etal:15,Zank.etal:15,leRoux.etal:16,Zhao.etal:19b}: the interaction of the  shocks with current sheets can generate magnetic islands thereby energizing particles. However, owing to the reasons outlined above, for the STA 2008DOY069 event current sheets are not a dominant effect in the particle acceleration.  

In summary, we have carried out test-particle simulations for protons accelerated at a laminar highly-oblique shock propagating into a medium with an embedded pre-existing turbulence to explain data for the {\it STEREO} quasi-perpendicular fast reverse shock of a stream interaction region 2008, DOY 069 (March $9^{th}$, UT: 19:50). We have shown that the main features of PADs and of intensity profiles of suprathermal protons ($10 -40$ keV and $84.1-92.7$ keV) measured within $10$ (downstream) $16$ (upstream) minutes from the shock are reproduced with excellent accuracy. The upstream bi-modal structure of PAD, due to bi-directional streaming of nearly scatter-free protons, can be hardly accounted for by shock-reflected ions scenarios, and its smearing-out further upstream, emerges clearly and consistently with earlier works \citep[e.g.,][]{Erdos.Balogh:94}. A systematic search of spikes at IP shocks is therefore needed to clarify whether or not PAD bi-modality is more likely to emerge at high-obliquity shocks as early models suggest \citep{Decker:83} or is also found at shocks with small obliquity propagating into highly turbulent SW. The prospect of measuring with {\it Parker Solar Probe} and {\it Solar Orbiter} a larger number of weak events close to the Sun with a very steep momentum spectrum (not many high energy particles) as compared to $1$ AU will help shed light on the role of $10^2 - 10^4$ km-scale of the magnetic structure in early-on phase of the particles acceleration at shocks.

\section*{Acknowledgements}

Helpful comments from the anonymous referee are acknowledged. The work of FF was supported, in part, by  NASA under Grants NNX15AJ71G and 80NSSC18K1213 and by NSF under grant 1850774. JG acknowledges support from NASA's Parker Solar Probe Mission (contract NNN06AA01C), 80NSSC18K1213 and NSF grant 1735422. FF thanks Drs. Wimmer-Schweingruber and Heber for clarifications on {\it PLASTIC} and {\it SEPT} instruments, Dr. Yang for sharing {\it PLASTIC} data and Dr. L. Jian for comments on the manuscript. Computational resources supporting this work were provided by the Hipas supercomputer cluster at the University of Arizona. The {\it STEREO/SEPT} project is supported under Grant 50 OC 1702 by the German Bundesministerium für Wirtschaft through the Deutsches Zentrum für Luft- und Raumfahrt (DLR); the {\it STEREO/PLASTIC} by NASA grant 80NSSC20K0431. We acknowledge data providers(s), J. Luhmann at UCB/SSL and CDAWeb for providing visualisation tool for the high-resolution MAG time series.

\section*{Data availability}

The {\it STEREO/SEPT} data underlying this article are publicly available at http://www2.physik.uni-kiel.de/stereo/index.php?doc=data. The 1/8 seconds resolution {\it IMPACT/MAG} data are publicly available at https://cdaweb.gsfc.nasa.gov/cgi-bin/eval3.cgi. The numerical code will be shared on reasonable request to the corresponding author.




\bibliographystyle{mnras}
\bibliography{ffraschetti} 








\bsp	
\label{lastpage}
\end{document}